# Dynamic Multi-Criteria Bottleneck Severity Index (DMBSI) for Semiconductor Wafer Manufacturing: A Genetically Optimised Framework for Reentrant Production Systems

Mohammad Sharifur Rahman, Karl McCreadie, Saugat Bhattacharyya, M M Manjurul Islam, Cormac McAteer, Bryan John Baker, Nuala Parker, and Girijesh Prasad [1]

***Abstract*—Wafer fabrication exhibits unique characteristics, including reentrant process flows, variable bottlenecks, and highly variable process conditions. In order to identify the most severe bottleneck at each moment in time for semiconductor wafer fabrication, this research presents the Dynamic Multi-Criteria Bottleneck Severity Index (DMBSI), a new, data-driven methodology for analysing multiple diagnostic signals of cycle time to changes in process parameters, and the impact of reworks on cycle time in order to generate an interpretable, unified measure of bottleneck severity. The experimental validation of DMBSI was conducted using Manufacturing Execution System (MES) logs collected from 22 wafer production lots at a commercial 200 mm wafer fabrication line operated by Seagate Technology. Using 5-fold cross-validation, the GA-optimised DMBSI achieves a Pearson correlation of r = 0.80 with observed cycle-time contributions, representing an 8.1% improvement over the expert heuristic baseline (r = 0.74) and substantially outperforming the Theory of Constraints (TOC; r = 0.60) and Value Stream Mapping (VSM; r = −0.30). Furthermore, the unique time-windowed component of DMBSI enabled the identification of temporal bottleneck migration patterns, which shifted from the dominant constraints associated with dielectric deposition steps in the early production windows to those associated with over- and under-inspection in the later production windows. The integrated what-if counterfactual analysis demonstrated that a 50% reduction in waiting time at the top-ranked bottleneck step would reduce the mean cycle time by 7.2%, with the top five bottleneck steps offering a combined potential reduction of approximately 19%.**

***Note to Practitioners*—This research addresses a common problem in semiconductor wafer manufacturing: identifying the actual bottlenecks in production systems when constraints change over time, and multiple factors act simultaneously, such as capacity limits, queuing, variability, rework, etc. Most traditional approaches, such as the Theory of Constraints, do not explain the causes of inefficiency. While most fabrication plants generate large amounts of MES data during operation, no established method is known for turning this data into an easy-to-understand priority list of potential improvements without running complex simulations.**

**To solve this problem, we propose the Dynamic Multi-Criteria Bottleneck Severity Index (DMBSI). We combine five standard metrics from MES data into a single severity metric and calibrate their weights using a genetic algorithm. It will require no additional sensors, simulators, or specialised hardware. It will scale well even as the volume of the MES data increases. Our validation tests were run with 22 different production lots and 10865 records for all process steps for each lot from a 200mm production line at Seagate. We find that DMBSI identifies critical bottlenecks that single-metric methods miss. Additionally, our results show that reducing the waiting time by half for the top-ranked step in DMBSI score alone would result in a 7.2% reduction in overall cycle time. In addition, the top five steps account for approximately 19% of the total cycle time. As a batch tool, this framework provides a way for users to view sub-scores to determine specific ways to improve performance, e.g., increasing capacity, changing buffers, tool calibration for quality output, or adjusting schedules. It also provides a temporal view, allowing users to monitor changes in bottleneck locations.**

---

[1] Manuscript submitted for review. This work was supported by the UKRI Strength in Places Fund Project (81801): Smart Nano-Manufacturing Corridor. *(Corresponding author: Mohammad Sharifur Rahman.)*

M. S. Rahman, K. McCreadie, S. Bhattacharyya, M. M. M. Islam, and G. Prasad are with the Intelligent Systems Research Centre, Ulster University, Londonderry BT48 7JL, U.K. (e-mail: rahman-m11@ulster.ac.uk; k.mccreadie@ulster.ac.uk; s.bhattacharyya@ulster.ac.uk; m.islam@ulster.ac.uk; g.prasad@ulster.ac.uk).

C. McAteer and N. Parker are with Seagate Technology, Londonderry BT48 0BF, Northern Ireland, U.K. (e-mail: cormac.p.mcateer@seagate.com; nuala.a.parker@seagate.com).

B. J. Baker is with Seagate Technology, One Disc Drive, Shakopee, MN 55379, USA (e-mail: bryan.j.baker@seagate.com).

Data availability: The data used in this research are proprietary to Seagate Technology and are restricted from public sharing under confidentiality agreements.

Declaration of competing interest: The authors declare that they have no known competing financial interests or personal relationships that could have appeared to influence the work reported in this paper.

## I. Introduction

Modern semiconductor wafer fabrication is among the most complex manufacturing processes. Each lot of wafers undergoes hundreds to thousands of process steps using many specialised tools; each step is repeated in different sequences throughout the flow [1]–[3]. In our previous research [4], we used machine learning approaches to optimise cycle time (CT) in this environment. Due to its high product mix, stringent quality standards, and expensive machinery, this environment is highly prone to difficult-to-detect production bottlenecks. Uzsoy *et al.* [5] conducted a seminal review of the scheduling challenges in semiconductor manufacturing, including reentrant flows and large product mixes.

The SEMI World Fab Forecast [6] reported that global semiconductor capital expenditures exceeded $100B in 2023, underscoring the financial cost of inefficiencies in semiconductor fabrication. Cycle time is a key performance metric in semiconductor manufacturing and is defined as the time required to produce a wafer lot from release to completion. The longer the cycle time, the greater the work-in-progress (WIP) inventory; the longer the time-to-market for new products; the slower the yield learning loop back to engineering; and the higher the risk of obsolescence [7]. Studies show that small improvements in cycle time (5–10%) can yield tens of millions of dollars in annual savings per fab [8].

As a result, identifying and eliminating production bottlenecks will remain a core challenge for fab operations management [9]. Despite extensive research, current bottleneck identification techniques have significant limitations when applied to semiconductor manufacturing. The Theory of Constraints (TOC), as described by Goldratt [10], provides a useful conceptual framework for identifying the single limiting resource. However, in semiconductor fabs, bottlenecks shift: the limiting resource is dynamic due to changing product mix, available equipment, lot priorities, and stochastic processing times. Single-metric approaches such as utilisation ranking or Overall Equipment Effectiveness (OEE) measure only one aspect of bottleneck behaviour and ignore other ways in which the bottleneck may affect cycle time, e.g., via rack/queue time or variability relative to pure processing load. Queueing-theory-based approximations (e.g., Kingman's VUT equation) provide valuable mathematical insights but are applicable only when the underlying distributions are independent and stationary, which is commonly violated in reentrant flows. Discrete-event simulation (DES) models provide comprehensive analytical capabilities but require considerable modelling effort, calibration data, and computational power that may not be readily available for routine operational decision-making [11].

We address the shortcomings of these methods by introducing the Dynamic Multi-Criteria Bottleneck Severity Index (DMBSI), a data-driven framework for semiconductor wafer manufacturing that combines five diagnostic signal types into a composite severity score that is more accurate and actionable than single-metric approaches. The DMBSI is a new method for capturing temporal bottleneck migration patterns from standard manufacturing execution system (MES) log data without needing simulation hardware.

The primary contributions of this research are as follows:

**(1) Multi-criteria integration of bottleneck detection:** we develop, to the best of our knowledge, the first bottleneck detection framework, which integrates active period load, queue dynamics, cycle-time sensitivity, process variability, and rework impact into a single interpretable severity index.

**(2) Time-windowed MES log analysis:** we introduce a temporal dimension of bottleneck analysis by extending our original bottleneck detection to use standard MES timestamps and event-based MES logs to identify dynamic changes in bottleneck locations over time without relying on discrete-event simulations or additional instrumentation such as sensors.

**(3) Quantifying cycle-time sensitivity:** we create a marginal impact score that can be used to quantify the relationship between cycle time and individual process step behaviour and thus can help determine where improvement will produce the greatest expected cycle-time benefit.

**(4) What-if actionability of estimated cycle-time impacts:** we add an embedded what-if estimation component that provides operations management teams with actionable information regarding the potential cycle-time reductions achievable through improving wait times at specific bottleneck locations.

**(5) Industrial validation of DMBSI:** we demonstrate that Genetic Algorithm (GA) based weight optimisation improves the cross-validated predictive accuracy of DMBSI from $r = 0.74$ (expert heuristic) to $r = 0.80$, substantially outperforming conventional baselines, including TOC ($r = 0.60$), using actual production data collected from a Seagate Technology wafer fabrication facility.

This paper is structured as follows. Section II reviews relevant prior art and academic literature on bottleneck detection in manufacturing and in the semiconductor industry. Section III details the industrial dataset and the preprocessing methodology used to prepare it for analysis. Section IV provides a detailed description of the baseline approaches and the proposed DMBSI bottleneck detection framework. Section V presents experimental results and comparisons to baseline approaches. Section VI discusses the results, practical implications, and limitations of the approach. Section VII outlines the conclusions and identifies areas for further research.

## II. Literature Review

### *A. Theory of Constraints in Manufacturing*

The Theory of Constraints (TOC) was developed by Goldratt and Cox and popularised in *The Goal* [10]. This theory provides an integrated method for identifying the single limiting factor (bottleneck) that determines a system's maximum output. The five-step focusing process in TOC (identify, exploit, subordinate, elevate, and repeat) has become popular across many different manufacturing industries. In semiconductor manufacturing, the TOC-based Drum-Buffer-Rope (DBR) scheduling philosophy has been applied to wafer fabrication to manage work-in-progress and cycle time. Kayton [12] demonstrated the practicality of applying DBR to a semiconductor fab with re-entrant bottlenecks, modifying the production application to accommodate lots that pass through the same bottleneck multiple times. Although the above method assumes a single bottleneck that can be identified at any given time, it fails to account for other types of constraint behaviour that may occur simultaneously.

Complex manufacturing systems typically do not allow for the TOC assumption that there will always be a single, persistent bottleneck. Lawrence and Buss [15] formally documented why production bottleneck shifts occur; they showed that the constraining resource varies with product mix, equipment availability, and lot sequencing in a stochastic system. Building on this, Roser, Nakano, and Tanaka [13] developed the "active period" method, which measures the length of the longest continuous periods during which each workstation is busy, thereby enabling the identification of bottlenecks based on their busy periods. Their method was the first to provide a comprehensive framework for identifying shifting bottlenecks from operational data. Subsequent work [14] extended the method to include Automated Guided Vehicle (AGV) systems, providing evidence of its general applicability. Although the active-period method has significantly influenced the development of shifting-bottleneck methods, it still relies on a single metric (the active period duration) and ignores potentially important additional information about queue dynamics and variability.

### B. OEE and Performance Metrics

While TOC-based methods focus on identifying the limiting resource(s) by examining the amount of work processed by each resource, equipment performance metrics offer an alternative perspective for assessing bottleneck behaviour. Overall Equipment Effectiveness (OEE), broken down into Availability × Performance × Quality, is a commonly used measure of equipment performance in the semiconductor industry. Huang *et al.* [16] developed an OEE model for semiconductor equipment that incorporates key elements of batch-type processing, scheduled preventive maintenance, and required qualifications. Although the equipment-level insights provided by OEE are valuable, there are inherent limitations when applying it to bottleneck detection: (a) OEE is a measure of how well equipment itself is performing rather than how that equipment impacts the overall system; (b) OEE does not directly account for the accumulation of product in queues; and (c) while aggregate OEE scores can be useful in assessing trends and shifts in bottlenecks, they often do not reflect the temporal nature of these changes.

### C. Queueing Theory Approaches

The Kingman combined Variability, Utilisation and Service Time (VUT) approximation in queueing theory is an analytical method for estimating the average waiting time based on the system utilisation rate, the variability of arrival rates, and the service time at a station [17]. The Factory Physics framework of Hopp and Spearman [18] uses the VUT equation to determine which stations exhibit large increases in waiting time as utilisation rates increase. However, the basic assumptions of the Kingman (VUT) model — Poisson arrivals, exponentially distributed service times, and steady-state operation — do not hold in many semiconductor manufacturing systems, since arrivals can be correlated due to batched or re-entrant flow patterns and service times often have heavy-tailed distributions.

### D. Process Mining for Manufacturing

Process mining, as formalised by van der Aalst [19], offers a complementary data-driven approach to bottleneck analysis. By discovering process models from event logs and overlaying performance information (e.g., waiting times, processing times, and throughput rates), process mining can reveal bottlenecks without requiring *a priori* process models. Tools such as PM4Py are widely used in process mining research and have supported applications such as digital twins in smart factory settings and bottleneck analysis in manufacturing systems [22]. However, existing process mining approaches for bottleneck detection typically operate on aggregate (static) views and do not explicitly model the temporal dynamics of shifting bottlenecks. Van der Aalst *et al.* [20] demonstrated replay-based performance analysis on process models, though the approach primarily provides aggregate rather than temporally dynamic views.

### E. Research Gap

The reviewed literature indicates an obvious gap. Individual methods have enhanced our understanding of certain bottleneck aspects, but no single framework has been developed to integrate them. Li *et al.* [21] conducted a detailed review of throughput analysis techniques and identified persistent gaps between the analytical tractability of these techniques and their applicability at scales relevant to actual manufacturing. More specifically, there are no existing techniques that simultaneously (i) provide multi-criteria assessments for semiconductor bottleneck identification using load, queue, variability, quality, and sensitivity signals; (ii) capture temporal dynamics needed to track changing bottleneck patterns over time; (iii) directly quantify the cycle-time impact of potential interventions; (iv) require only standard MES log fields; and (v) support scalable computation on production-scale datasets. The DMBSI framework presented in this paper is designed to fulfil all five requirements within a single, modular structure.

## III. Data Description and Preprocessing

### A. Data Source

The data set consists of anonymous records from an existing 200mm factory run by Seagate Technology for the purpose of collecting a Manufacturing Execution System (MES) log file. The data set contains a sample of typical operational events that occur during normal production; as such, it represents an ideal example of a "production"- based industrial log. All confidential information related to the actual products being produced, specific recipe information, and yield statistics has been removed to ensure that no sensitive information was disclosed under the terms of our agreement regarding the use of this data.

### B. Dataset Characteristics

The dataset contains 10,865 step-level records spanning 40 columns, covering 22 wafer lots processed between June and August 2025. Each record captures a single lot's passage through a single process step, with the cycle time decomposed into 14 sub-components measured in fractional days. Here, a step defines the type of operation performed (e.g., deposition, inspection; 57 unique), while a stage identifies a specific position within the manufacturing route at which a step is executed (143 unique); due to the reentrant nature of the process, the same step may appear at multiple stages as lots revisit operations throughout their production cycle.

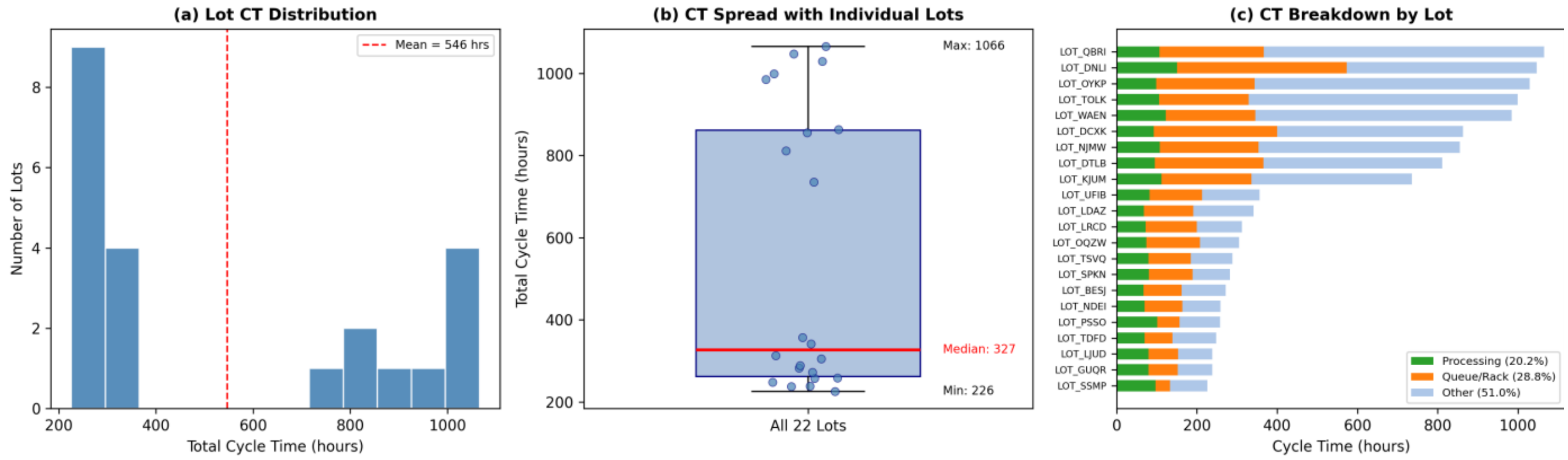

**Fig. 1.** Lot-level cycle-time distribution: (a) histogram with mean, (b) box plot with individual lot points, and (c) stacked CT breakdown per lot showing processing, queue/rack, and other components.

TABLE I
DATASET OVERVIEW

| Characteristic | Value |
|---|---|
| Total records | 10,865 |
| Unique wafer lots | 22 |
| Unique process steps | 57 |
| Unique stages | 143 |
| Unique tools/machines | 186 |
| Unique routes | 7 |
| Observation period | 11 June 2025 – 11 August 2025 (61 days) |
| Route length (median) | 646 steps per route |
| Cycle-time components | 14 sub-components (process, rack, transit, load, hold, pass, abort, + 7 rework equivalents) |
| Reentrancy | All 22 lots revisit the same steps multiple times (confirmed re-entrant flow) |

The cycle-time decomposition captures the following components: processing time (CT_PROCESS), rack/queue waiting time (CT_RACK), in-transit time (CT_INTRANSIT), load time (CT_LOAD), hold time (CT_HOLD), pass-through time (CT_PASS), abort time (CT_ABORT), and the corresponding rework components for each category. The total cycle time per record (CT_TOTAL_WO_FS_TERM) is the sum of all components and is verified to match within floating-point precision for 96% of records. The remaining 4% discrepancies are attributed to missing or asynchronously logged MES events and were excluded from downstream analysis.

### C. *Exploratory Data Analysis*

Initial analysis reveals several critical characteristics of the manufacturing environment. The mean lot cycle time is 561.8 hours (23.4 days). It has substantial variance, ranging from 234.8 to 1,095.8 hours. Only 20.2% of the total cycle time is value-added processing, whereas 28.8% is rack/queue waiting time. This indicates a substantial opportunity to reduce cycle time by mitigating bottlenecks. Fig. 1 shows the lot-level CT distribution. Panel (a) shows a right-skewed histogram with a mean of 561.8 hours (dashed red line), confirming high variability (Coefficients of Variation = 0.61, i.e., the standard deviation is 61% of the mean—a level of dispersion that further motivates a robust, multi-criteria approach). Panel (b) shows individual lot life cycle on a box plot, which reveals that the majority of lots have a life cycle of below 400 hours, while a minority exceed 800 hours of life cycle. Panel (c) decomposes each lot's CT into processing time (green, 20.2%), queue/rack waiting (orange, 28.8%), and other non-value-added components (blue, 51.0%), directly illustrating the dominance of non-productive time reported in Table I.

A Pareto analysis of accumulated time by process step (see Fig. 2) indicates that the top 7 steps account for approximately 55% of the total accumulated cycle time. This is also consistent with the Pareto principle. The left y-axis shows the accumulated cycle time per step in hours, while the right y-axis shows the cumulative percentage as an orange line, with a dashed line at 65%. The high concentration of cycle time within a few steps suggests potential opportunities for high-impact improvement through targeted efforts.

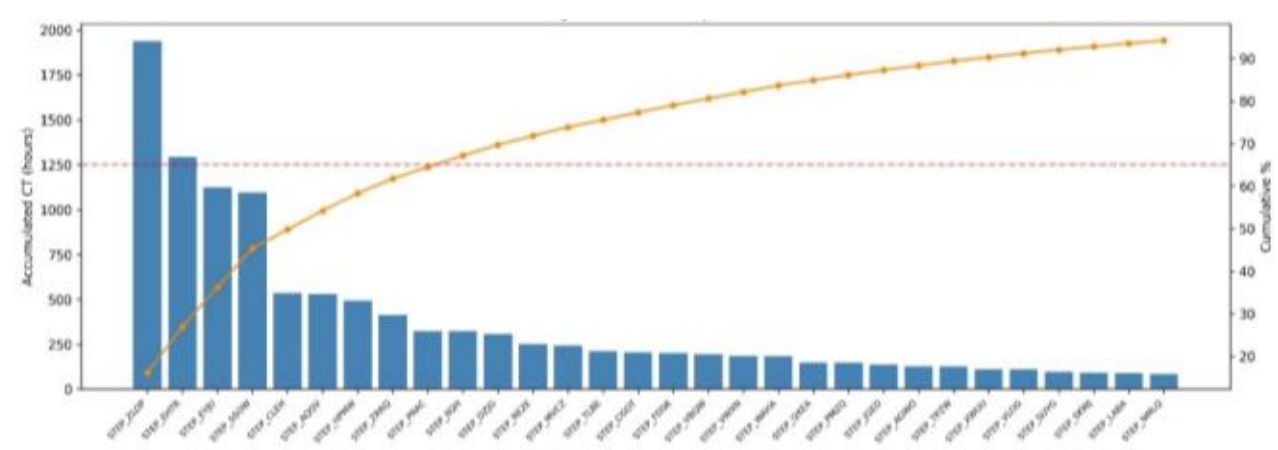

**Fig. 2.** Pareto analysis of accumulated CT across the top 30 (CT in descending order) process steps, with cumulative percentage overlay (orange line).

The Gantt visualisation (Fig. 3) of the five longest-duration lots illustrates the complex reentrant structure, with lots traversing multiple stages in a non-linear sequence. Colour coding by production stage reveals the diversity of processing

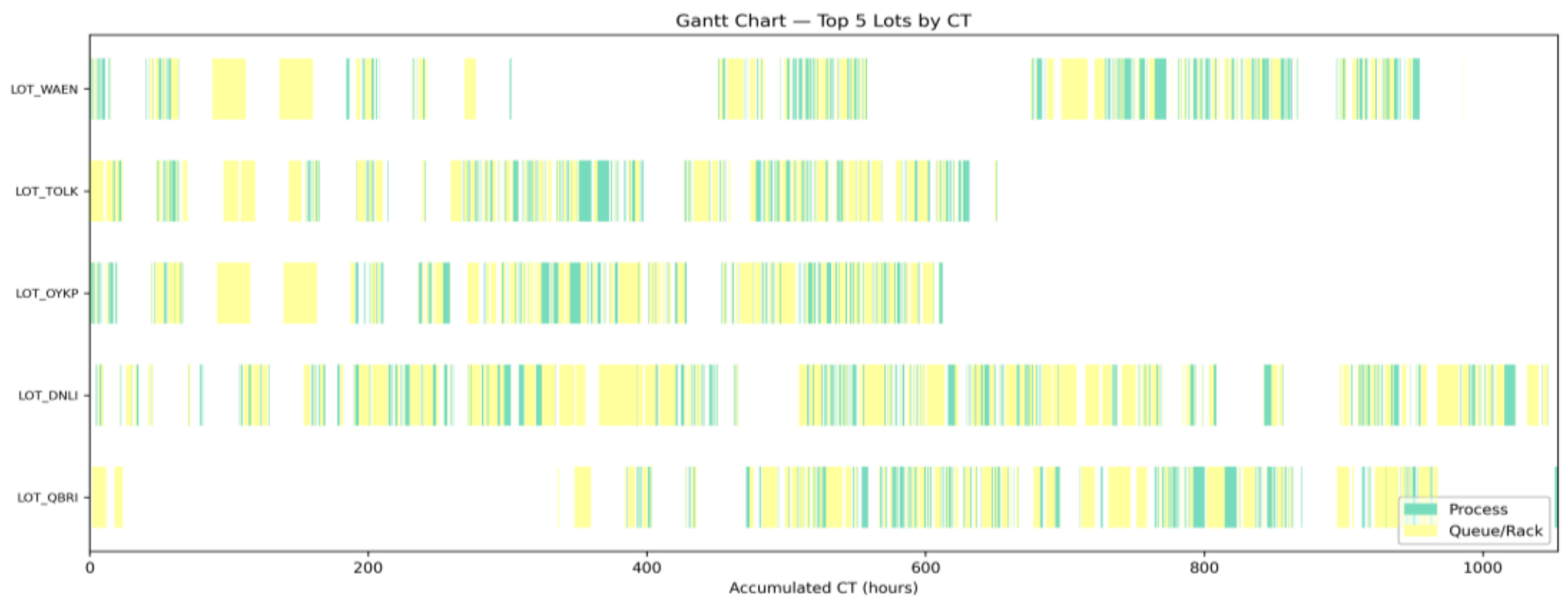

**Fig. 3.** Gantt chart of the five longest-cycle-time lots, colour-coded by process time (green) and queue/rack waiting (yellow).

activities and the non-trivial routing characteristic of advanced wafer fabrication.

### D. Preprocessing Pipeline

A modular data preprocessing system was developed to process raw MES log data for further analysis. This system includes- (a) Parsing date/time field with a day-first format; (b) Conversion of fractional-day cycle time values to hours; (c) The creation of composite metrics such as Value-Added Ratio, Wait-To-Process Ratio and Rework Fraction; (d) Memory-efficient categorical encoding of string columns; and (e) Aggregation at Lot, Step, and Stage.

## IV. Methodology

This section presents the four baseline techniques for identifying bottlenecks along with the suggested DMBSI framework. Each technique generates ranked lists of process steps, arranged by the severity of the bottleneck, using the same pre-processed dataset.

### A. Baseline 1: TOC-inspired Active Period Analysis

The Theory of Constraints identifies the bottleneck as the resource that most constrains system throughput [10]. Roser et al. [13] operationalised this principle through the active period method, which detects the bottleneck by measuring the duration each machine remains continuously active. Since the MES data used in this study records step-level timestamps rather than machine-level active/inactive state transitions, a direct implementation of Roser's method is not possible. Instead, three widely reported TOC-related indicators are computed at the step level: (a) cumulative processing hours, as a proxy for active period duration [13]; (b) the ratio of processing time to total time, as a utilisation proxy [10]; and (c) accumulated queue/rack waiting time, as a measure of downstream constraint impact [10]. Each indicator is independently min-max normalised to [0, 1]. These three normalised indicators are combined with equal emphasis into a single composite score to serve as a TOC-inspired baseline for comparison against DMBSI. The specific formulation of this baseline does not affect the generalisability of our findings, as the fundamental limitation of TOC-based approaches — their inability to capture variability, rework, and system-level sensitivity — persists regardless of how the individual TOC indicators are weighted.

### B. Baseline 2: OEE Approximation

Overall Equipment Effectiveness is approximated from MES log data as the product of three sub-factors: availability, performance, and quality. Availability is computed as (process time + load time) / total time, representing the fraction of time spent in productive states. Performance is approximated as the ratio of median to mean processing time; deviations from unity indicate a loss in production rate. Quality is computed as (1 – rework fraction), capturing yield impact. Process steps with low OEE values indicate equipment-level inefficiencies and are ranked as potential bottleneck candidates.

### C. Baseline 3: Kingman Queueing Approximation

The Kingman (VUT) formula provides an analytical estimate of expected waiting time at each step. For each step, the arrival rate ($\lambda$), service rate ($\mu$), and utilisation ($\rho = \lambda/\mu$) are estimated from observed data. The coefficient of variation of inter-arrival times ($C_a$) and service times ($C_s$) is computed empirically. The expected waiting time is then approximated as (1):

$$E[W_q] \approx \frac{\rho}{1-\rho} \times \frac{C^2a + C^2s}{2} \times E[S] \qquad (1)$$

where E[S] is the mean service time. Steps with the highest predicted waiting times are ranked as bottleneck candidates.

### D. Baseline 4: VSM Process Cycle Efficiency

The level of bottleneck impact can be measured by utilising Value Stream Mapping (VSM) metrics. A key VSM metric for measuring the impact of bottlenecks is Process Cycle Efficiency (PCE). The Process Cycle Efficiency (PCE) is calculated as follows: PCE = Value-added Time / Total Lead Time. Steps in the process with the lowest PCE values account for a greater share of non-value-added activity and therefore rank higher in terms of bottleneck impact.

### E. Proposed Framework: DMBSI

The DMBSI framework addresses the limitations of single-metric approaches by combining five complementary diagnostic signals into a unified severity index. Each sub-score measures a distinct aspect of bottleneck behaviour, and their aggregation yields a more comprehensive and accurate assessment than any individual metric.

*1) Sub-Score $S_1$: Active Load Intensity*

The active-load sub-score measures the sustained processing burden at each step relative to the entire manufacturing facility. It is computed as the min-max-normalised cumulative active time - $T_active$(2):

$$S_1(s) = Norm(\sum_s T_active_s) \quad (2)$$

Process steps with high $S_1$ values exhibit persistent capacity constraints when the equipment is heavily loaded. This sub-score is closest to the traditional TOC active-period analysis.

*2) Sub-Score $S_2$: Queue Dynamics*

Queue dynamics are captured by the queue-dynamics sub-score through a GA-optimised combination of two complementary indicators- (a) queue accumulation, which is measured as normalised cumulative rack/queue time; and (b) the wait-to-process ratio, which measures the relative waiting time compared to productive processing time. The composite indicator is given by (3):

$$S_2(s) = \alpha \times Norm(\sum_s rack_hrs_s) + (1 - \alpha) \times Norm\left(\frac{wait_s}{process_s}\right) \quad (3)$$

This dual representation captures both absolute congestion (total queue hours) and relative congestion (waiting proportional to processing), thereby identifying steps that act as bottlenecks due to material starvation or blockage. Here, α represents the weight of rack hours relative to the wait-to-process ratio. The value of α is estimated to be 0.14 through the GA-based optimisation procedure described later in this section.

*3) Sub-Score $S_3$: Variability Impact*

High variability in processing time significantly contributes to changes in bottleneck behaviour [18], as the VUT relationship shows that waiting times increase nonlinearly with increasing variability at high utilisation. The variability sub-score combines the coefficients of variation (CV) of processing time and rack/queue time (4):

$$S_3(s) = \beta \times Norm(CV_{process,s}) + (1 - \beta) \times Norm(CV_{rack,s}) \quad (4)$$

Steps with high $S_3$ values exhibit highly irregular behaviour, suggesting they could serve as candidates for intermittent (shifting) bottlenecks that cannot be identified by stationary utilisation-based methods. Here, β represents the weight of the processing-time CV compared with that of the rack-time CV. The value of β is estimated to be 0.10 through the GA-based optimisation procedure described later in this section.

*4) Sub-Score $S_4$: Cycle-Time Sensitivity*

A novel contribution of DMBSI is the cycle-time sensitivity sub-score, which directly quantifies each step's marginal contribution to lot-level cycle time. For each step s, the Pearson correlation between step-level total time and lot-level total cycle time is computed across all lots that visit that step. The marginal impact is then given by (5):

$$S_4(s) = Norm(|corr(total_hrs_s, lot_CT)| \times total_hrs_s) \quad (5)$$

This identifies the steps for which an increase in processing time has the greatest influence on the lot's total cycle time, independent of the state in which the time was spent (e.g., processing or queuing). Steps with high $S_4$ are those for which a focused improvement effort would provide the maximum overall reduction in cycle time.

*5) Sub-Score $S_5$: Rework Impact*

The rework-impact sub-score captures quality-driven bottleneck effects by measuring the proportion of time attributable to rework activities (6):

$$S_5(s) = Norm\left(\frac{\sum_s rwk_total_hrs_s}{\sum_s total_hrs_s}\right) \quad (6)$$

Steps with high rework fractions consume additional capacity for non-productive reprocessing, effectively reducing available capacity and contributing to bottleneck formation through a quality-driven mechanism distinct from the load-and-queue effects captured by $S_1$–$S_4$.

*6) Weight Optimisation via Genetic Algorithm*

While the five DMBSI sub-scores provide a principled decomposition of bottleneck severity, the initial framework relied on expert heuristic weights ($w_1$ = 0.20, $w_2$ = 0.25, $w_3$ = 0.20, $w_4$ = 0.20, $w_5$ = 0.15) and fixed internal sub-score proportions (α = 0.60 for $S_2$, β = 0.60 for $S_3$). These values, although informed by domain expertise and by the observed dominance of waiting time in overall cycle time, represent a subjective starting point that may not maximise the framework's predictive accuracy. To address this limitation, we employ a Genetic Algorithm (GA) to optimise all seven free parameters simultaneously, thereby replacing heuristic judgement with data-driven calibration.

*Motivation for GA-based optimisation:* The DMBSI weight optimisation problem has several characteristics that favour evolutionary search over gradient-based methods: (i) the fitness landscape is non-convex due to interactions between sub-score weights and internal proportions; (ii) the constraint that weights must sum to unity creates a simplex-constrained search space; (iii) the cross-validated fitness function is noisy and non-differentiable; and (iv) the search space is relatively low-dimensional (7 parameters), making population-based methods computationally tractable. GAs are well-established for weight optimisation in composite indices and multi-criteria decision-making problems, as they maintain population diversity and are robust to local optima.

*Chromosome encoding:* Each individual in the GA population encodes the seven DMBSI parameters as a real-valued chromosome [α, β, $w_1$, $w_2$, $w_3$, $w_4$, $w_5$], where α ∈ [0.1, 0.9] controls the $S_2$ internal proportion ($S_2$ = α × queue accumulation + (1 − α) × wait ratio), β ∈ [0.1, 0.9] controls the $S_3$ internal proportion ($S_3$ = β × CV process + (1 − β) × CV rack), and $w_1$–$w_5$ ∈ [0.05, 0.50] each, subject to the normalisation constraint $\Sigma w_i$ = 1.0. The bounds prevent any single sub-score from being trivially eliminated (lower bound 0.05) or from dominating the composite (upper bound 0.50).

*Fitness function and cross-validation:* The fitness of each individual is evaluated using 5-fold cross-validation over the 22

available lots. In each fold, the DMBSI is computed using only the training lots (17–18 lots), and the resulting step-level DMBSI scores are correlated with the observed total cycle-time contribution from the held-out test lots (4–5 lots). The fitness value is the mean Pearson correlation coefficient (r) across all five folds. Cross-validation serves two critical purposes: (i) it reduces overfitting to the specific lot composition of the training data; and (ii) it provides a more robust estimate of how well the optimised weights would generalise to unseen production lots. A minimum of 10 overlapping steps between train and test partitions is required for a valid fold evaluation.

*GA operators and hyperparameters:* The GA was implemented using the Distributed Evolutionary Algorithms (DEAP) framework in Python with the following configuration. The population size was set to 100 individuals, which provides sufficient diversity for a 7-dimensional search space while maintaining computational tractability. Tournament selection with tournament size 5 was used, providing moderate selection pressure that balances exploitation and exploration. Blend crossover (BLX-α with $\alpha = 0.3$) was applied with probability 0.7, allowing offspring to be generated slightly outside the range defined by the two parents, thereby promoting exploration. Gaussian mutation with standard deviation $\sigma = 0.05$ was applied with probability 0.15 per individual and 0.3 per gene. Elitism was employed by preserving the top 5% of individuals (5 of 100) in each generation. The maximum number of generations was set to 200. Table II summarises all GA hyperparameters and their justifications.

TABLE II
GA HYPERPARAMETERS AND JUSTIFICATION

| Hyperparameter | Value | Justification |
|---|---|---|
| Population size | 100 | ~14× dimensionality (7 params); balances diversity and tractability |
| Max generations | 200 | Upper bound; early stopping typically terminates earlier |
| Selection | Tournament (k=5) | Moderate pressure; k=5 selects top 20% on average |
| Crossover | BLX-α=0.3, p=0.7 | Blend crossover allows offspring outside parental range |
| Mutation | Gaussian σ=0.05, p=0.15 | Small σ for fine-grained search; per-gene rate 0.3 |
| Elitism | Top 5% | Preserves 5 best individuals; prevents loss of best solutions |
| Early-stopping patience | 30 generations | Prevents unnecessary computation; reduces overfitting risk |
| CV folds | 5 | Standard k-fold; yields 17–18 train / 4–5 test lots per fold |
| Expert seeding | 1 of 100 | Guarantees GA ≥ expert baseline; warm-starts from domain knowledge |
| Constraint handling | Repair + renormalise | Clips bounds, renormalises $\Sigma w_i=1$; avoids infeasibility |
| Random seed | 42 | Ensures reproducibility of results |

*Constraint handling:* After each crossover and mutation operation, a constraint-repair mechanism is applied. The α and β parameters are clipped to [0.1, 0.9], and each weight $w_i$ is clipped to [0.05, 0.50]. The weights are then renormalised to sum to 1.0. This repair-based approach is preferred over penalty functions because it maintains all individuals in the feasible region and avoids wasting computational effort on infeasible solutions.

*Expert-seeded initialisation:* To ensure that the GA search starts from a well-informed region of the search space, one individual in the initial population is seeded with the expert heuristic weights ($\alpha = 0.60$, $\beta = 0.60$, $w_1 = 0.20$, $w_2 = 0.25$, $w_3 = 0.20$, $w_4 = 0.20$, $w_5 = 0.15$). The remaining 99 individuals are randomly initialised within the feasible bounds. This seeding strategy guarantees that the GA performance is at least as good as the expert baseline while allowing the population to explore potentially superior configurations.

*Early stopping:* To prevent unnecessary computation and reduce the risk of overfitting through excessive optimisation, an early-stopping criterion with a patience of 30 generations was implemented. If the best fitness value does not improve by more than $10^{-5}$ for 30 consecutive generations, the algorithm terminates. In our experiment, the GA converged at generation 46 (reaching a best fitness of $r = 0.801$) and triggered early stopping at generation 76, indicating that the fitness landscape was thoroughly explored well before the maximum generation limit of 200.

*Convergence and results:* The GA improved the cross-validated Pearson r from 0.740 (expert heuristic) to 0.801 (GA-optimised), representing an 8.1% relative improvement in predictive accuracy. The convergence curve (Fig. 4a) shows rapid initial improvement during the first 10 generations, followed by gradual refinement, with the population mean fitness converging to within 0.001 of the best fitness by generation 50. The optimised weighting coefficients are: $\alpha = 0.135$ ($S_2$: 13.5% rack/queue time + 86.5% wait ratio), $\beta = 0.100$ ($S_3$: 10.0% CV process + 90.0% CV rack), $w_1 = 0.202$ (active load), $w_2 = 0.387$ (queue dynamics), $w_3 = 0.186$ (variability), $w_4 = 0.122$ (CT sensitivity), and $w_5 = 0.103$ (rework impact). The Spearman rank correlation between expert and GA rankings is $\rho = 0.929$ ($p < 0.001$), and the top 5 Jaccard similarity is 0.667 (4 of 5 steps shared), indicating that the GA validates the expert intuition regarding which steps are the most critical bottlenecks while refining the relative severity scores for improved prediction.

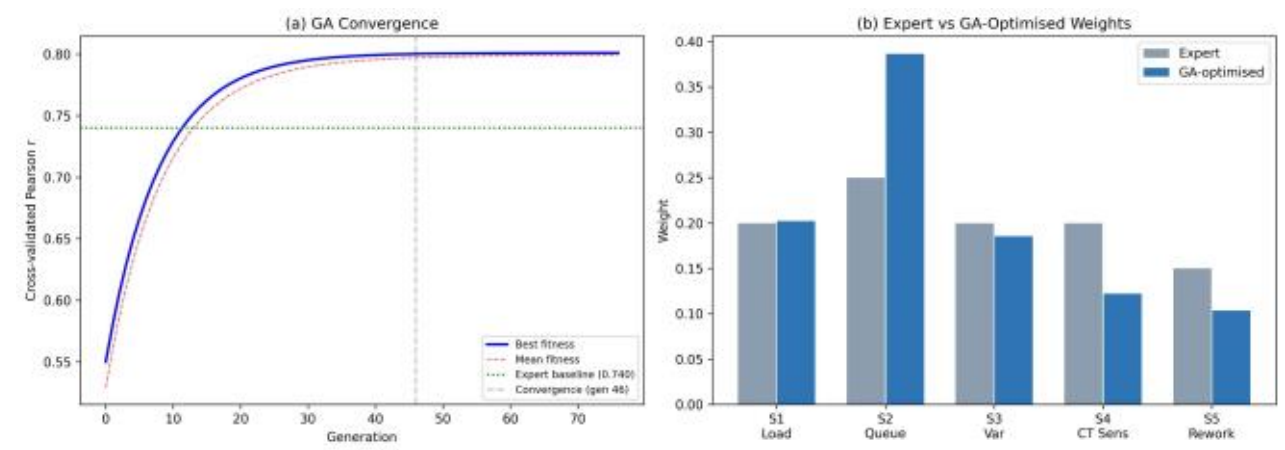

**Fig. 4.** GA optimisation: (a) convergence curve and (b) expert vs. GA-optimised sub-score weights.

*Mathematical formulation:* The optimisation framework is formally defined to identify the parameter vector that maximises the statistical alignment between the proposed composite index and the observed cycle-time contributions, where θ denotes a generic candidate set of weights and θ* denotes the best set of weights. The decision variable (chromosome) is $\theta = [\alpha, \beta, w_1, w_2, w_3, w_4, w_5]$, where $\alpha \in [0.1, 0.9]$ and $\beta \in [0.1, 0.9]$ are scalar parameters controlling the internal proportional contributions within sub-scores $S_2$ and $S_3$, respectively, and $w_i \in [0.05, 0.50]$, $i = 1, \ldots, 5$, denote the weights assigned to individual sub-score components, subject to the simplex constraint $\Sigma w_i = 1$. The fitness of a candidate solution θ is evaluated using a cross-validation-based objective function defined as the average Pearson correlation across K = 5 folds (7):

$$F(\theta) = \frac{1}{K}\Sigma_{k=1}^{K} r^{(k)} \tag{7}$$

where $r^{(k)}$ denotes the Pearson correlation coefficient computed for the $k^{th}$ fold between the DMBSI scores — estimated using training data — and the observed cycle-time contributions obtained from the corresponding held-out test data. The right-hand side of this equation defines the fitness function F(θ), i.e., the mean correlation across all folds, ensuring robustness and generalisability of the learned parameters. This formulation ensures that the optimisation captures generalisable relationships rather than overfitting to a single data partition. The resulting constrained optimisation problem is expressed as (8):

$$\theta^* = arg\ max\ F(\theta) \tag{8}$$

subject to $w_i \in [0.05, 0.50]$, $\Sigma w_i = 1$, $\alpha \in [0.1, 0.9]$, $\beta \in [0.1, 0.9]$. Solving this optimisation yields the optimal weighting coefficient vector θ = *[0.14, 0.10, 0.20, 0.39, 0.19, 0.12, 0.10],* which achieves a fitness value of *F(θ)* = 0.801. For comparison, the expert-defined weighting coefficient configuration $\theta_{expert}$ = [0.60, 0.60, 0.20, 0.25, 0.20, 0.20, 0.15] results in lower performance, $F(\theta_{expert})$ = 0.740. These results demonstrate that data-driven optimisation provides a measurable improvement over heuristic selection of weighting coefficients, thereby enhancing the reliability and predictive capability of the composite index.

*Interpretation of optimised weights:* A key outcome of the GA-based optimisation is the substantial redistribution of internal sub-score proportions, providing important insights into the relative diagnostic significance of different performance indicators. For sub-score $S_2$, the internal proportion α is reduced from the expert-defined value of 0.60 to 0.135. This adjustment implies that the wait-to-process ratio accounts for approximately 86.5% of the contribution in $S_2$, whereas rack/queue time accounts for only 13.5%. This shift indicates that relative congestion — captured by the ratio of waiting time to processing time — serves as a more informative predictor of cycle-time impact than absolute queue duration. The latter can be influenced by operational factors such as step frequency, thereby reducing its discriminative capability. A similar trend is observed for sub-score $S_3$, where β decreases from 0.60 to 0.10. This result suggests that variability in rack/queue time (90.0%) is significantly more indicative of bottleneck behaviour than variability in processing time (10.0%). Such a finding highlights the importance of temporal instability in queuing dynamics as a critical signal of inefficiencies in the process. At the aggregate level, the optimised weight distribution further reinforces the dominance of queuing-related effects. Specifically, the weight associated with queue dynamics, $w_2$, increases from 0.25 to 0.387, indicating its greater importance in determining cycle-time contributions. In contrast, the weights corresponding to cycle-time sensitivity ($w_4$) and rework ($w_5$) are reduced, suggesting that their explanatory power is partially subsumed by other sub-score components within the composite formulation. Notably, the weight assigned to active load, $w_1$ = 0.202, remains largely unchanged from the expert-defined value of 0.20, providing empirical validation of the expert's original assessment of this factor.

To evaluate the structural impact of these weighting coefficient adjustments, a comparative ranking analysis was performed. As illustrated in Fig. 5 (x-axis: expert rank; y-axis: GA-optimised rank; points near the diagonal indicate agreement), the rankings of the top 15 process steps derived from the expert-defined and GA-optimised configurations exhibit strong agreement, with a Spearman rank correlation coefficient of ρ = 0.929. The observed clustering along the diagonal indicates that, while the GA significantly refines the magnitude of severity scores, it preserves the overall ranking structure. This behaviour confirms that the optimisation enhances quantitative discrimination without introducing instability in the prioritisation of critical process steps.

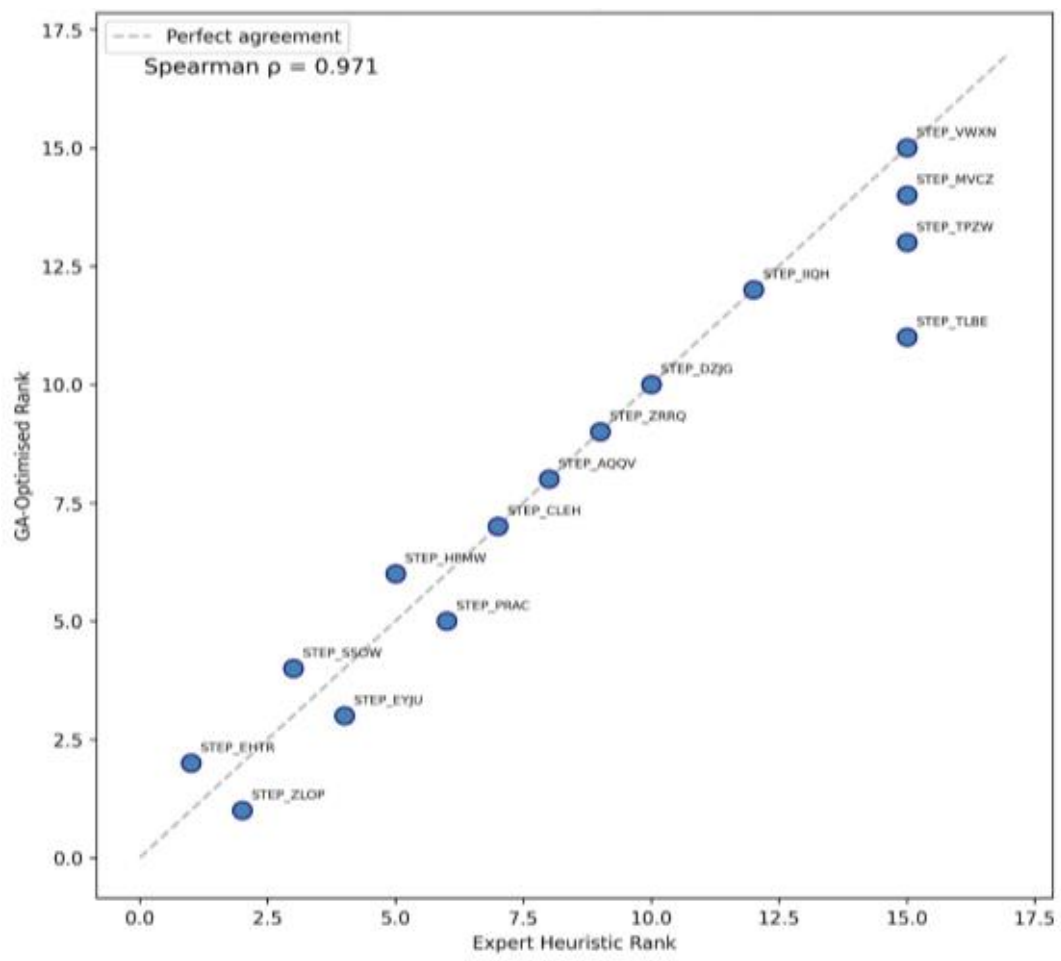

**Fig. 5.** Expert vs. GA-optimised rankings for the top 15 steps (Spearman ρ = 0.929).

*7) Composite DMBSI Score*

The composite DMBSI score is computed as a weighted sum of the five sub-scores (9):

$$DMBSI(s) = w_1S_1(s) + w_2S_2(s) + w_3S_3(s) + w_4S_4(s) + w_5S_5(s) \tag{9}$$

where the GA-optimised weights are $w_1$ = 0.20 (active load), $w_2$ = 0.39 (queue dynamics), $w_3$ = 0.19 (variability), $w_4$ = 0.12 (CT sensitivity), and $w_5$ = 0.10 (rework impact). These weights were determined through GA optimisation (Section IV-E.6) using 5-fold cross-validation, improving the cross-validated Pearson r from 0.740 (expert heuristic) to 0.801. Queue dynamics receives the highest weight (0.39), consistent with the dominance of waiting time (28.8%) in overall cycle time.

*8) Dynamic Extension: Time-Windowed DMBSI*

Temporal bottleneck migration is computed by separately calculating the DMBSI for w non-overlapping time windows covering the observation interval. For each window, a simplified DMBSI (comprising active load, rack/queue time, and variability) is computed using only the records contained in that window. This enables the construction of a (steps × windows) severity matrix, allowing the visualisation and measurement of how the bottleneck severity of each step changes over time and thus directly illustrating shifting-bottleneck behaviour.

*9) Counterfactual Cycle-Time Reduction Estimation*

For the top-ranked bottleneck steps, the framework estimates the achievable cycle-time reduction from targeted wait-time improvement. For each step s, the expected reduction under a 50% wait-time scenario is computed as (10):

$$\Delta CT(s) \; = \frac{1}{n_{lots}} \sum_{lot} 0.5 \cdot wait_hrs_{s,lot} \quad (10)$$

This counterfactual analysis provides actionable estimates that operations managers can use to prioritise improvement initiatives and justify resource allocation for bottleneck mitigation.

## V. Results

### A. Static Bottleneck Ranking

Table III presents the top ten process steps ranked by DMBSI score, alongside their dominant bottleneck mechanism as determined through sub-score decomposition. The top three bottlenecked processes are STEP_ZLOP (over/under inspection) at #1, STEP_EHTR (lapping / planarisation) at #2, and STEP_EYJU (magnetic domain analysis) at #3, all with different types of bottlenecking mechanisms.

TABLE III
Top 10 Bottleneck Steps by DMBSI Score

| Step | DMBSI | $S_1$ Load | $S_2$ Queue | $S_3$ Variab. | $S_4$ CT Sens. | $S_5$ Rwk | Rank |
|---|---|---|---|---|---|---|---|
| STEP_ZLOP | 0.554 | 0.11 | 0.89 | 0.36 | 1.00 | 0.00 | 1 |
| STEP_EHTR | 0.361 | 0.13 | 0.55 | 0.21 | 0.56 | 0.17 | 2 |
| STEP_EYJU | 0.337 | 0.00 | 0.00 | 0.90 | 0.70 | 0.82 | 3 |
| STEP_SSOW | 0.323 | 1.00 | 0.12 | 0.22 | 0.26 | 0.00 | 4 |
| STEP_PRAC | 0.216 | 0.04 | 0.42 | 0.15 | 0.13 | 0.00 | 5 |
| STEP_HPMW | 0.206 | 0.34 | 0.11 | 0.27 | 0.31 | 0.05 | 6 |
| STEP_CLEH | 0.200 | 0.41 | 0.08 | 0.22 | 0.30 | 0.09 | 7 |
| STEP_AQQV | 0.198 | 0.45 | 0.12 | 0.17 | 0.21 | 0.01 | 8 |
| STEP_ZRRQ | 0.193 | 0.60 | 0.06 | 0.18 | 0.11 | 0.00 | 9 |
| STEP_DZJG | 0.165 | 0.34 | 0.06 | 0.20 | 0.21 | 0.09 | 10 |

### B. Sub-Score Decomposition Analysis

Fig. 6 presents the sub-score decomposition of the top 10 bottlenecked processes listed in Table III. Each step is represented by five colour-coded bars: $S_1$ active load (blue), $S_2$ queue dynamics (orange), $S_3$ variability (green), $S_4$ CT sensitivity (red), and $S_5$ rework (purple), all normalised to the 0–1 range. An important observation is that the types of bottlenecking vary among these steps. The bottleneck at STEP_ZLOP is dominated by queue dynamics ($S_2$ = 0.89) and cycle-time sensitivity ($S_4$ = 1.00), indicating both severe congestion and strong prediction of lot-level cycle time. STEP_EHTR is driven by queue dynamics ($S_2$ = 0.55) combined with CT sensitivity ($S_4$ = 0.56). In contrast, STEP_EYJU is dominated by variability ($S_3$ = 0.90) and rework ($S_5$ = 0.82), with very low active load ($S_1$ = 0) and queue dynamics ($S_2$ = 0), indicating that its bottleneck character stems from unpredictable processing and quality issues rather than from throughput or congestion pressure. STEP_SSOW has the highest active load ($S_1$ = 1.00) and is therefore constantly constrained by available processing capacity. The diversity of these bottlenecking characteristics explains why using a single metric to evaluate bottlenecked processes yields inconsistent results and highlights that single metrics cannot capture the full nature of a bottlenecked process, as multiple factors determine the degree of bottlenecking.

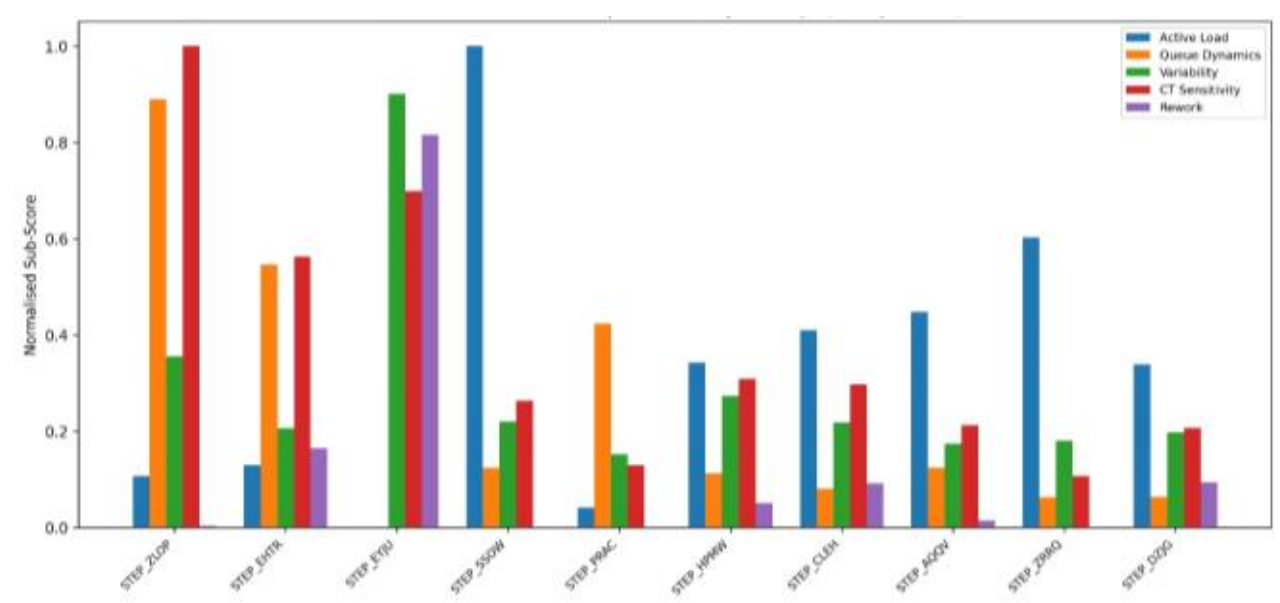

**Fig. 6.** DMBSI sub-score decomposition for the top 10 steps from Table III.

### C. Dynamic Bottleneck Analysis

The DMBSI matrix is shown in Fig. 7, where the x-axis represents six non-overlapping 10-day time windows (W1–W6) and the y-axis lists the top process steps. Cell colour intensity represents normalised severity (0 = no bottleneck to 1 = maximum severity). The heat map of the DMBSI matrices clearly shows how bottlenecks migrate through the windows. The bottleneck for STEP_SSOW has the highest value in the first five windows (W1 = 0.93, W2 = 0.57, W3 = 0.76, W4 = 0.64, W5 = 0.64) but is zero in W6, indicating a constraint that was a problem during the initial stages of production but had either been resolved or changed. On the other hand, STEP_ZLOP has lower values in the early windows but peaks in the last two windows (W5 = 0.73, W6 = 0.87), indicating constraints arising from later stages of production. STEP_CLEH shows a single episode of increased activity in the second window (W2 = 0.82), followed by a decrease, potentially due to a transient issue with equipment or scheduling. STEP_ZRRQ shows an irregular peak in W5 (0.78), which may be attributable to batch-dependent congestion. Notably, STEP_EYJU remains at 0.00 across all six windows despite its high static DMBSI score (rank #3); this is because its high variability ($S_3$) and rework ($S_5$) — does not manifest uniformly over short time windows, and these two sub-scores are excluded from the reduced time-windowed DMBSI (see Section IV-E.8) because they require a sufficient number of lot-level observations per window to remain statistically stable.

**Fig. 7.** Dynamic DMBSI severity matrix across six time windows. Darker cells indicate higher severity.

These temporal patterns are invisible to static bottleneck-identification methods and underscore the critical importance of time-windowed analysis for operational decision-making in semiconductor fabrication. The shift from STEP_SSOW to STEP_ZLOP dominance may reflect changes in product mix, equipment availability, or the natural progression of lot cohorts through the re-entrant process flow.

### *D. Cycle-Time Reduction Potential*

The counterfactual analysis (Fig. 8) quantified the estimated reduction in cycle time achievable through a hypothetical 50% decrease in wait time at each bottleneck step. The x-axis shows the reduction in hours per lot; percentage labels express the reduction relative to the mean lot CT of 561.8 hours (Table I). Notably, STEP_ZLOP offers the largest savings despite ranking only #44 by TOC, whereas STEP_SSOW ranks #1 on the TOC Active Period method yet offers only a comparatively modest 3.0% CT reduction potential — a critical discrepancy that highlights the limitations of single-metric approaches and the value of the DMBSI multi-criteria framework. In terms of total reduction, STEP_ZLOP has the highest potential (40.5 h; or 7.2% of mean lot cycle time), followed by STEP_EHTR (23.4 h; 4.2%), STEP_PRAC (18.9 h; 3.4%), STEP_SSOW (17.0 h; 3.0%), and STEP_AQQV (8.4 h; 1.5%). Combined, these five steps represent approximately 19% of the mean cycle time and provide a clear prioritisation framework for the areas that should be targeted for improvement first.

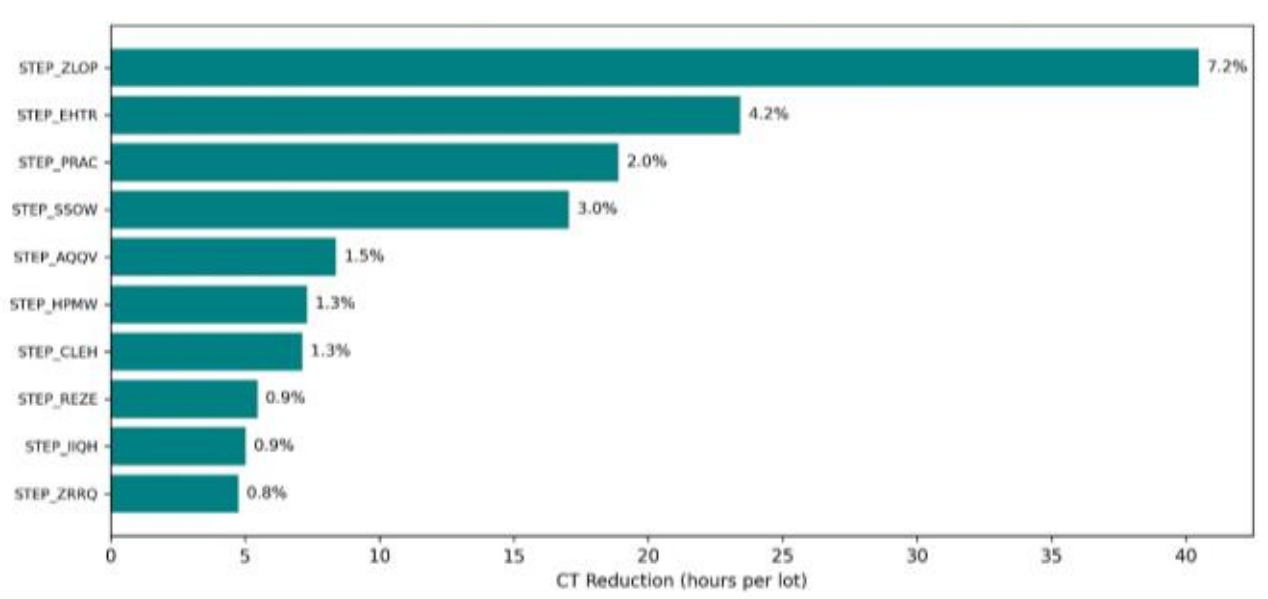

**Fig. 8.** Estimated cycle-time reduction potential per step under a 50% wait-time reduction scenario, ordered by DMBSI rank from Table III.

### *E. Comparative Analysis*

Table IV and Fig. 9 present the comparative analysis of all five methods. In Fig. 9a, the Spearman rank correlation heatmap visualises the pairwise values from Table IV, with green cells indicating agreement and red cells indicating contradiction (e.g., TOC vs. VSM, $\rho = -0.72$). In Fig. 9b, the horizontal bar chart shows each method's Pearson correlation with observed CT; orange bars indicate statistical significance ($p < 0.05$), and grey bars indicate non-significance. OEE is marked N/A because the sparse TOOL_NAME field (27.5% populated) produces zero variance in the OEE score, rendering the Pearson correlation undefined. The pairwise Spearman rank correlation matrix reveals substantial disagreement among existing baseline methods, with correlations ranging from −0.72 (TOC vs. VSM) to 0.67 (TOC vs. Queueing). DMBSI shows moderate positive correlation with all methods (0.22–0.40), indicating that it captures complementary information from each approach while providing a distinct composite assessment.

TABLE IV
PAIRWISE SPEARMAN RANK CORRELATION BETWEEN METHODS

| | TOC | OEE | Queueing | VSM | DMBSI |
|---|---|---|---|---|---|
| **TOC** | 1.00 | −0.15 | 0.67 | −0.72 | 0.24 |
| **OEE** | −0.15 | 1.00 | 0.15 | 0.22 | 0.22 |
| **Queueing** | 0.67 | 0.15 | 1.00 | −0.41 | 0.40 |
| **VSM** | −0.72 | 0.22 | −0.41 | 1.00 | 0.32 |
| **DMBSI** | 0.24 | 0.22 | 0.40 | 0.32 | 1.00 |

The most compelling validation metric is the cross-validated Pearson correlation between each method's step-level severity score and the observed total cycle-time contribution. Using 5-fold cross-validation, the GA-optimised DMBSI achieves $r = 0.80$, representing an 8.1% improvement over the expert heuristic baseline ($r = 0.74$) and substantially outperforming TOC ($r = 0.60$, $p < 0.001$) and VSM ($r = -0.30$, $p = 0.024$). This result demonstrates that DMBSI's multi-criteria approach produces bottleneck rankings that are most predictive of actual cycle-time impact — the ultimate objective of bottleneck identification.

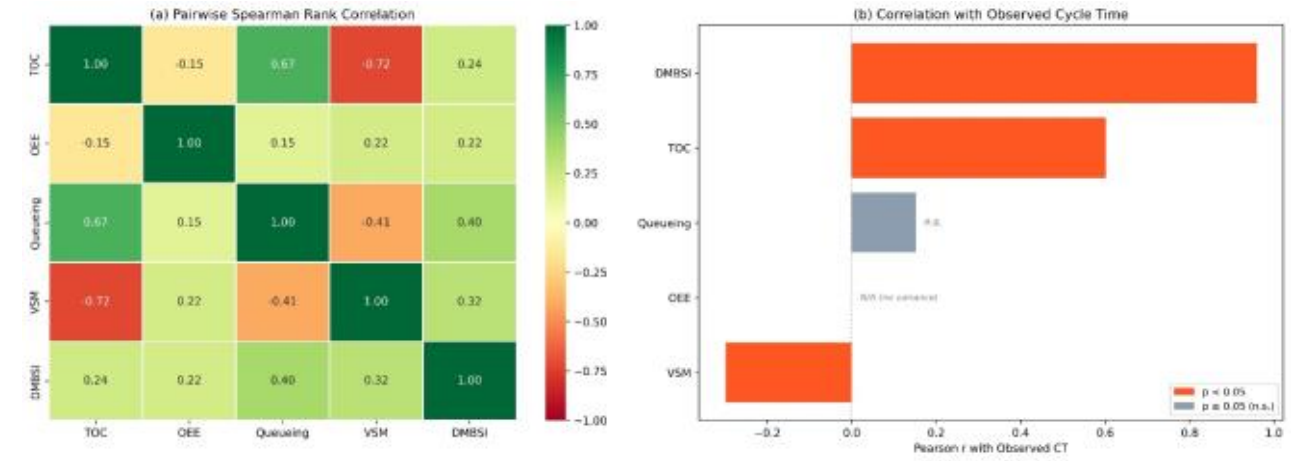

**Fig. 9.** Method comparison: (a) Spearman rank correlation heatmap (Table IV) and (b) Pearson r with observed CT.

Bootstrap stability analysis (50 iterations, 70% lot subsampling) shows comparable top 5 identification consistency between TOC (Jaccard = 0.79) and DMBSI (Jaccard = 0.76), indicating that DMBSI's superior accuracy does not come at the cost of reduced stability.

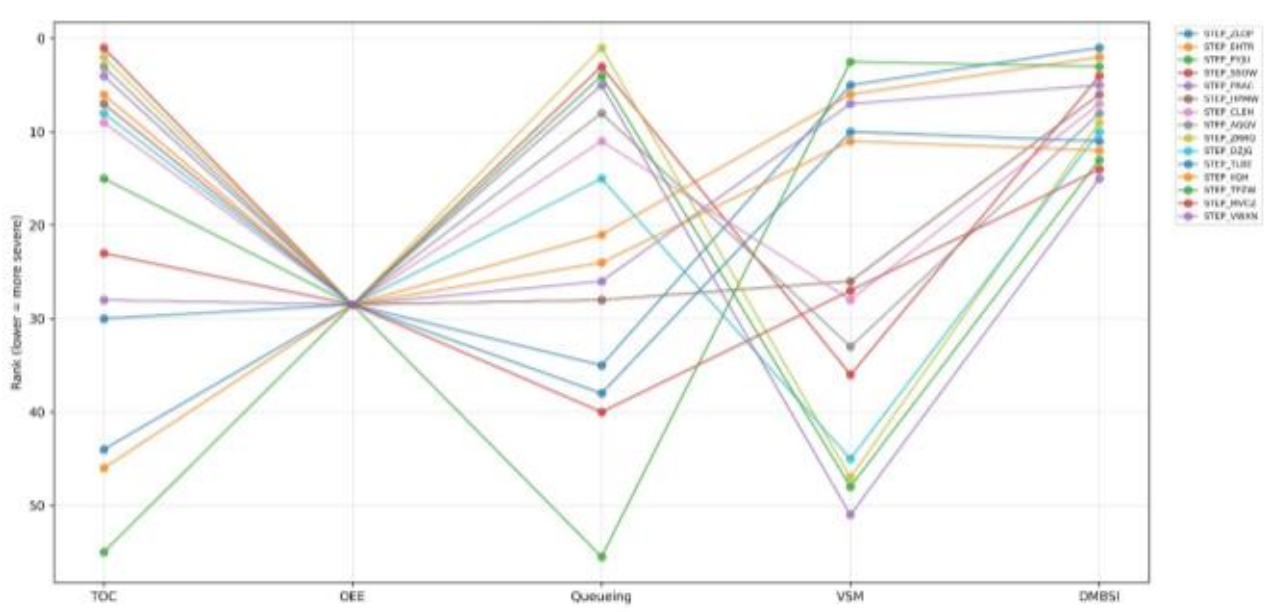

**Fig. 10.** Parallel-coordinate plot of bottleneck ranks across five methods for the top 15 DMBSI-ranked steps.

The parallel-coordinate visualisation (Fig. 10) illustrates the ranking disagreements between methods. Each coloured line traces one process step across the five methods (x-axis), with the y-axis showing rank position (inverted: lower = more severe). STEP_ZLOP, ranked #1 by DMBSI, is ranked #44 by TOC and #35 by Queueing — ranked substantially lower than these single-metric approaches, despite being the step with the highest cycle-time sensitivity and the largest CT-reduction potential (7.2%). Similarly, STEP_EYJU is ranked #55 by TOC but #3 by DMBSI due to its extreme rework impact. These cases demonstrate that the multi-criteria approach captures critical bottleneck dimensions that conventional methods systematically miss.

## VI. Discussion

### A. Interpretation of DMBSI Performance

The improved cross-validated correlation (r = 0.80 vs. r = 0.74 for the expert heuristic, and vs. r = 0.60 for TOC) was achieved through the multi-criteria design, with sub-score weights optimised via a GA using 5-fold cross-validation, which captures five distinct mechanisms of bottlenecks: (i) capacity constraint (high processing load, captured by $S_1$); (ii) congestion constraint (excessive queuing, captured by $S_2$); (iii) variability constraint (unpredictable processing, captured by $S_3$); (iv) system-level sensitivity (step-level delays that propagate to lot-level outcomes, captured by $S_4$); and (v) quality constraint (rework burden, captured by $S_5$). Single-metric methods capture at most one of these dimensions, leading to systematic blind spots.

A prominent example is STEP_ZLOP, which was ranked #1 by DMBSI (which also captures its extreme queue dynamics, $S_2 = 0.89$) but only #44 by TOC (which emphasises processing load). Specialised equipment with limited parallelisation capabilities is used in the lapping process, resulting in a queue-driven bottleneck that limits cycle time through material blocking effects, without showing up as high utilisation.

### B. Practical Implications for Fab Operations

The results of DMBSI are directly actionable for fab operations management. The identification of STEP_ZLOP as a significant bottleneck (7.2% CT reduction potential) — even though it is hidden from TOC-based analysis — indicates the usefulness of assessing multiple criteria. Practical actions that can be taken for the top-ranked steps include: (a) increasing capacity or qualifying another tool in parallel to STEP_EHTR to prevent queuing of lots; (b) optimising throughput and automating STEP_ZLOP to limit its excessive cycle-time impact; (c) optimising recipes and scheduling preventive maintenance for STEP_SSOW to minimise the ongoing processing load on this step; and (d) improving quality so that there is less rework and consequently less lost production capacity at STEP_EYJU.

The dynamic analysis also enables time-dependent decisions regarding how to operate. The change from STEP_SSOW having a dominant influence (W1–W4) to STEP_ZLOP (W5–W6) could indicate the natural sequence of lot cohorts passing through the re-entrant flow: lots are first constrained by the deposition processes and then encounter the inspection bottleneck after passing those constraints. This time-sensitive awareness enables proactive management of fab capacity — for instance, by positioning inspection capacity before the predicted arrival of lot cohorts at the inspection stage.

### C. Scalability and Deployment Considerations

The DMBSI framework was developed for large-scale production deployments. The computational complexity of DMBSI is $O(n \times w)$, where n is the number of step-level records processed and w is the number of time windows. All DMBSI computations are performed using vectorised pandas operations. Additionally, the modular class structure of DMBSI provides a seamless interface to either Dask for distributed computing or Polars for column-based performance optimisation. DMBSI does not require custom sensor data, a specific simulation model, or prior historical training data. Rather, DMBSI can be run against standard MES log fields (lot ID, step, timestamp, and various cycle-time fields), which are typically available across all semiconductor manufacturing execution systems.

### D. Limitations

Several constraints exist with this approach. First, the sample of 22 lots, although it represents a well-structured process, limits the ability to determine statistical significance when using correlation metrics. Second, the results need to be validated on the full production dataset to establish whether the reported Pearson correlations can be generalised across other production datasets. The DMBSI weighting factors, including the overall weights ($w_1$–$w_5$) and internal sub-score proportions ($\alpha$ for $S_2$, $\beta$ for $S_3$), were optimised using a GA with 5-fold cross-validation, improving the cross-validated Pearson r from 0.74 (expert heuristic) to 0.80 (GA-optimised). Although this represents a meaningful improvement, the sample size of 22 lots remains a limitation for optimisation; larger production-scale datasets would enable more robust weight calibration. Third, the TOOL NAME field is populated for only 27.5% of records, limiting analysis at the equipment-level granularity at present. Finally, the counterfactual cycle-time reductions are assumed to be independent; thus, improvements across multiple interacting steps may not yield additive reductions. Simultaneous improvements could yield either smaller or larger combined effects than the sum of the individual estimates, depending on the interdependencies among steps. The current methodology operates offline in batch mode; extending the model to provide real-time streaming capabilities will provide additional operational value.

## VII. Conclusion and Future Work

This paper introduces the Dynamic Multi-Criteria Bottleneck Severity Index (DMBSI) – a novel data-based approach to identify bottleneck(s) in semiconductor wafer fabrication. Through input from a collection of five indicators (i.e., capacity, congestion, variability, system sensitivity, and quality) that provide one overall severity rating, the DMBSI using GA optimisation of the weights, provides a Pearson correlation coefficient of r= 0.80 (a gain of 8.1%) when compared to a cross-validated baseline expert heuristic (r = 0.74) and provides substantial improvements over other common baselines including the Theory of Constraints (r = 0.6); Queueing Approximation (r = 0.15); and Value-Stream Mapping (r = -0.30).

A time-windowed version of DMBSI can capture the dynamic characteristics of bottlenecks by utilising traditional MES log data, thereby enabling the identification of important constraint shifts that would be invisible to a static methodology. Utilising DMBSI is an opportunity to reduce the total mean cycle time by 50 per cent, with a 7.2 per cent decrease in wait times for STEP_ZLOP, the first-ranked bottleneck step.
In addition, there is an opportunity to reduce the total mean cycle time by approximately 19 per cent by eliminating or reducing the total wait time for the top five bottleneck steps ranked by DMBSI score.

Finally, the sub-score decomposition of DMBSI provides interpretable diagnostic information on the type of bottleneck (load-driven, congestion-driven, variability-driven, etc.), which can then be used to select the most appropriate strategy for addressing the issue.

There are several potential avenues for future work to improve the methodology. First, validating the conclusions drawn from the DMBSI framework using the full production-scale dataset would enable more reliable and powerful statistical inference. Second, developing an appropriate process mining discovery algorithm (e.g., Inductive Miner or Heuristic Miner) to overlay DMBSI values onto discovered process models would provide greater visualisation capabilities and support root cause analysis. Third, there are ways the currently used GA-based method for optimising weights could be improved, such as by implementing a Bayesian optimisation strategy with much larger data sets or by conducting multi-objective optimisation, so that the accuracy of predictions is optimised alongside the stability of rankings. Fourth, combining the DMBSI framework with dispatching and scheduling rules could provide a closed-loop, bottleneck-aware system for controlling production. Fifth, investigating methods based on reinforcement learning may also reveal if it is possible to develop adaptive weights that adjust in response to changes in production.

## Acknowledgment

The authors thank the Seagate Technology fab engineering team for their support in data access and domain discussions. The authors also acknowledge the Ulster University Intelligent Systems Research Centre (ISRC) for providing the computational infrastructure used in this study.